\pdfoutput=1
\documentclass[aps,pra,twocolumn,10pt]{revtex4-1}
\usepackage[utf8]{inputenc}
\usepackage{fourier}
\usepackage{amsmath,graphicx,microtype,bm}
\usepackage[cal=boondoxo,frak=boondox]{mathalfa}
\usepackage[colorlinks=true,citecolor=blue,urlcolor=blue]{hyperref}

\begin{document}

\title{Image-potential-induced spin-orbit interaction in one-dimensional electron systems}

\author{Yasha Gindikin and Vladimir A.\ Sablikov}

\affiliation{Kotel'nikov Institute of Radio Engineering and Electronics,
Russian Academy of Sciences, Fryazino, Moscow District, 141190, Russia}

\begin{abstract}
We study the spin-orbit interaction effects in a one-dimensional electron system that result from the image charges in a nearby metallic gate. The nontrivial property of the image-potential-induced spin-orbit interaction (iSOI) is that it directly depends on the electron density because of which a positive feedback arises between the electron density and the iSOI magnitude. As a result, the system becomes unstable against the density fluctuations under certain conditions. In addition, the iSOI contributes to the electron-electron interaction giving rise to strong changes in electron correlations and collective excitation spectra. We trace the evolution of the spectrum of the collective excitations and their spin-charge structures with the change in the iSOI parameter. One out of two collective modes softens as the iSOI amplitude grows to become unstable at its critical value. Interestingly, this mode evolves from a pure spin excitation to a pure charge one. At the critical point its velocity turns to zero together with the charge stiffness.
\end{abstract}

\maketitle
\section{Introduction}

The Rashba spin-orbit interaction (RSOI) in low-dimensional systems arises because of a structure inversion asymmetry, which results from an external electric field acting on electrons in addition to the crystalline field. The RSOI plays a central role in such areas as the generation, manipulation and detection of spin, topological states, Majorana fermions, low-dimensional materials with Dirac-type spectra and even cold-atom systems (for a recent review see Ref.~\cite{manchon2015new}).

The RSOI is described by the Rashba Hamiltonian~\cite{winkler2003spin}
\begin{equation}
\label{rham}
	H_\mathrm{RSOI} = \alpha (\bm{\mathfrak{E}} \times \mathbf{k}) \bm{\sigma}\,,
\end{equation}
where $\bm{\mathfrak{E}}$ is an external electric field, which is usually considered as a given value. By tuning the field $\mathfrak{E}$, one can gain control over the RSOI parameter $\alpha_R = \alpha \mathfrak{E}$. This is important for the spin manipulation by electrical means.

\begin{figure}[htb]
	\includegraphics[width=0.9\linewidth]{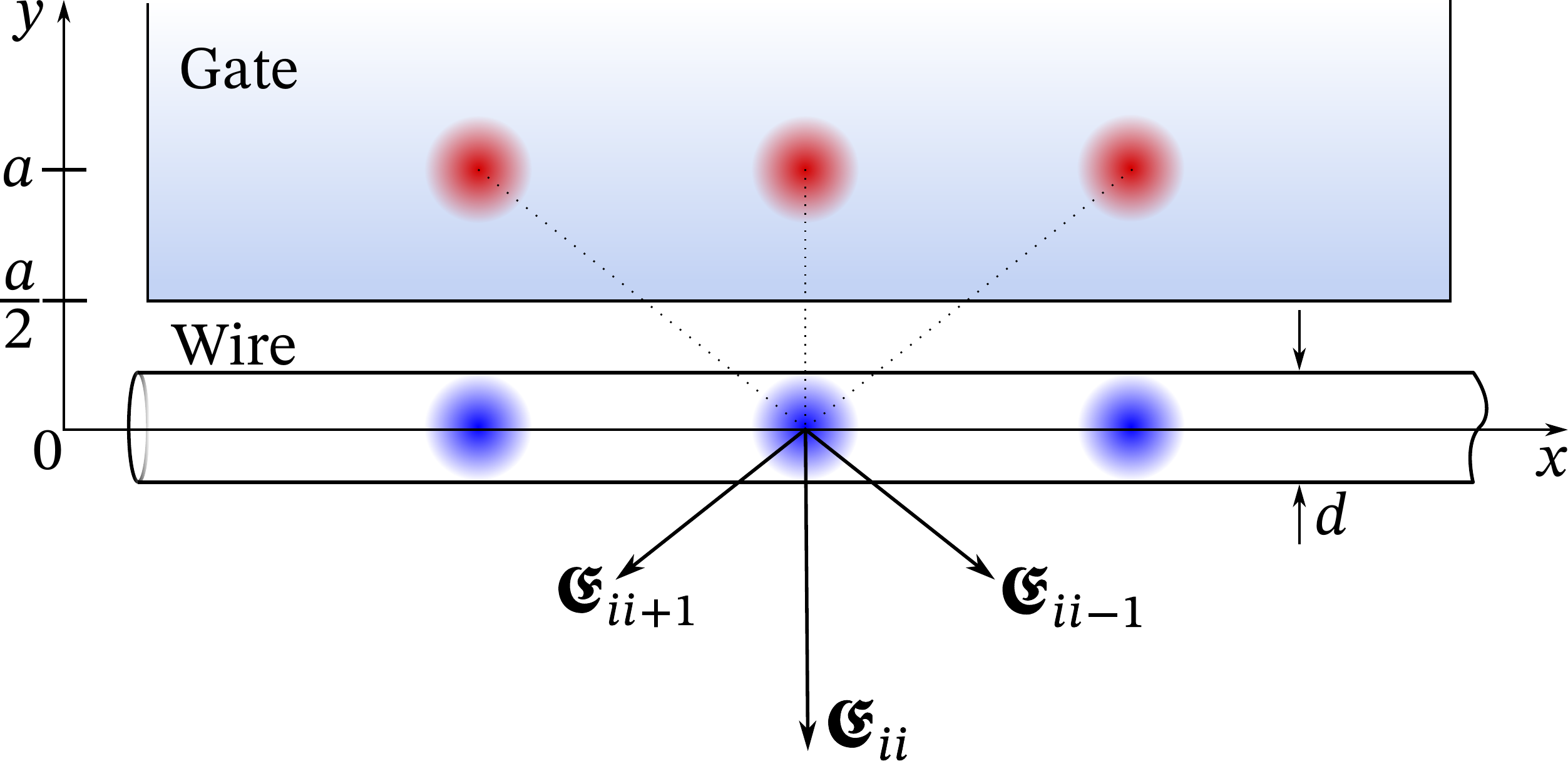}
	\caption{The schematic of a one-dimensional electron system with image charges induced on a gate. The arrows show the electric fields acting on electrons from their own image charges and from the images of neighboring electrons.}
\label{fig1}
\end{figure}

In the present paper we consider a principally different situation where the structure symmetry is broken by a metallic gate placed in close proximity to the electronic system and coupled to it by the Coulomb forces. This situation is close to the experiments where the electron system under investigation is placed directly on a conductive gate~\cite{Bachtold1317}. In this case the RSOI can arise even without any potential applied to the gate thanks to the image charges electric field as shown in Fig.~\ref{fig1}. This field is strong enough in the vicinity of the interface. One may therefore expect strong effects due to the image-potential-induced spin-orbit interaction (iSOI). The presence of the iSOI recently was confirmed by several experiments where the spin-orbit splitting was observed in the surface electron states formed by the image potential on the Au(001) surface~\cite{PhysRevB.94.115412} and at the graphene/Ir(111) interface~\cite{PhysRevLett.115.046801}. The values of $\alpha_R$ measured in these experiments agree well with the calculations performed by McLaughlan \textit{et al}.~\cite{0953-8984-16-39-017}. 

A novel and fascinating property of the iSOI is that $\alpha_R$ depends on the electron density. This dependence creates an efficient mechanism for density fluctuations to grow, which under certain circumstances can result in a dramatic transformation of the ground state. The mechanism is as follows. An electron density fluctuation induces an additional image charge and hence increases an electric field component normal to the gate surface. This enhances the iSOI parameter $\alpha_R$ and consequently lowers the electron energy within the fluctuation region, attracting there electrons from adjacent regions or reservoirs. Thus the density fluctuation once appeared starts to grow.

\section{Qualitative considerations}

Let us begin with a qualitative description of the process. To be specific, consider a single-mode quantum wire parallel to a metallic gate, separated by a distance of $a/2$ from the latter. Let us determine the electron density in the wire for the case of a fixed chemical potential $\mu$. For now, we restrict ourselves to a mean-field theory, assuming the electron density $n$ to be uniformly distributed.

The single-electron state energy reads as
\begin{equation}
\label{energy_functional}
	\varepsilon_{ks} = \frac{\hbar^2}{2m}[{(k + s\,k_\mathrm{so})}^2 - k_\mathrm{so}^2] + \mathfrak{v} \,n\,,
\end{equation}
where $k$ is the longitudinal wave vector and $s = \pm 1$ is the spin index. The Coulomb interaction energy is $\mathfrak{v} = \frac{2e^2}{\epsilon} \ln (a/d)$ with $d$ being the quantum wire diameter and $\epsilon$ as the dielectric constant. The iSOI wave vector is $k_\mathrm{so} = \alpha_R m/\hbar^2$. The iSOI parameter $\alpha_R=\alpha \mathfrak{E}_{\mathrel \bot}$ is proportional to the normal component of the electric field where the SOI constant $\alpha$ does not depend on the field. It is important that the field is determined by the electron density $\mathfrak{E}_{\mathrel \bot} = 2n e/\epsilon a$. Whence it follows that $k_\mathrm{so}=2en \alpha m/\hbar^2 \epsilon a$. The equation for the electron density is found by summing over the occupied states. Taking into account that there are two values of the Fermi momenta for each spin direction, $k_F^{(s)} = -s\,k_\mathrm{so} \pm {[k_\mathrm{so}^2 + 2m(\mu - \mathfrak{v}\,n )/\hbar^2]}^{1/2}$, we obtain an equation to determine $n$ at zero temperature,
\begin{equation}
\label{instab_n}
	n = \frac{2}{\pi}\sqrt{{\left(\frac{2\alpha me}{\hbar^2 \epsilon a}\right)}^2n^2 + \frac{2m}{\hbar^2}(\mu - \mathfrak{v} n)}\,.
\end{equation}

Its solutions are
\begin{equation}
\label{branches}
	n_{\pm}(\alpha^*) = n_0 \frac{-\mathfrak{v}^* \pm \sqrt{1 - {\alpha^*}^2 + {\mathfrak{v}^*}^2}}{1 - {\alpha^*}^2}\,,
\end{equation}
where $n_0 = \sqrt{8 m \mu}/ \pi \hbar$, $\mathfrak{v}^*=\mathfrak{v} \sqrt{2 m \mu^{-1}}/ \pi \hbar$, and $\alpha^* = 4 \alpha m e/ \pi \hbar^2 \epsilon a$ is a dimensionless iSOI parameter.

\begin{figure}[htb]
	\includegraphics[width=0.9\linewidth]{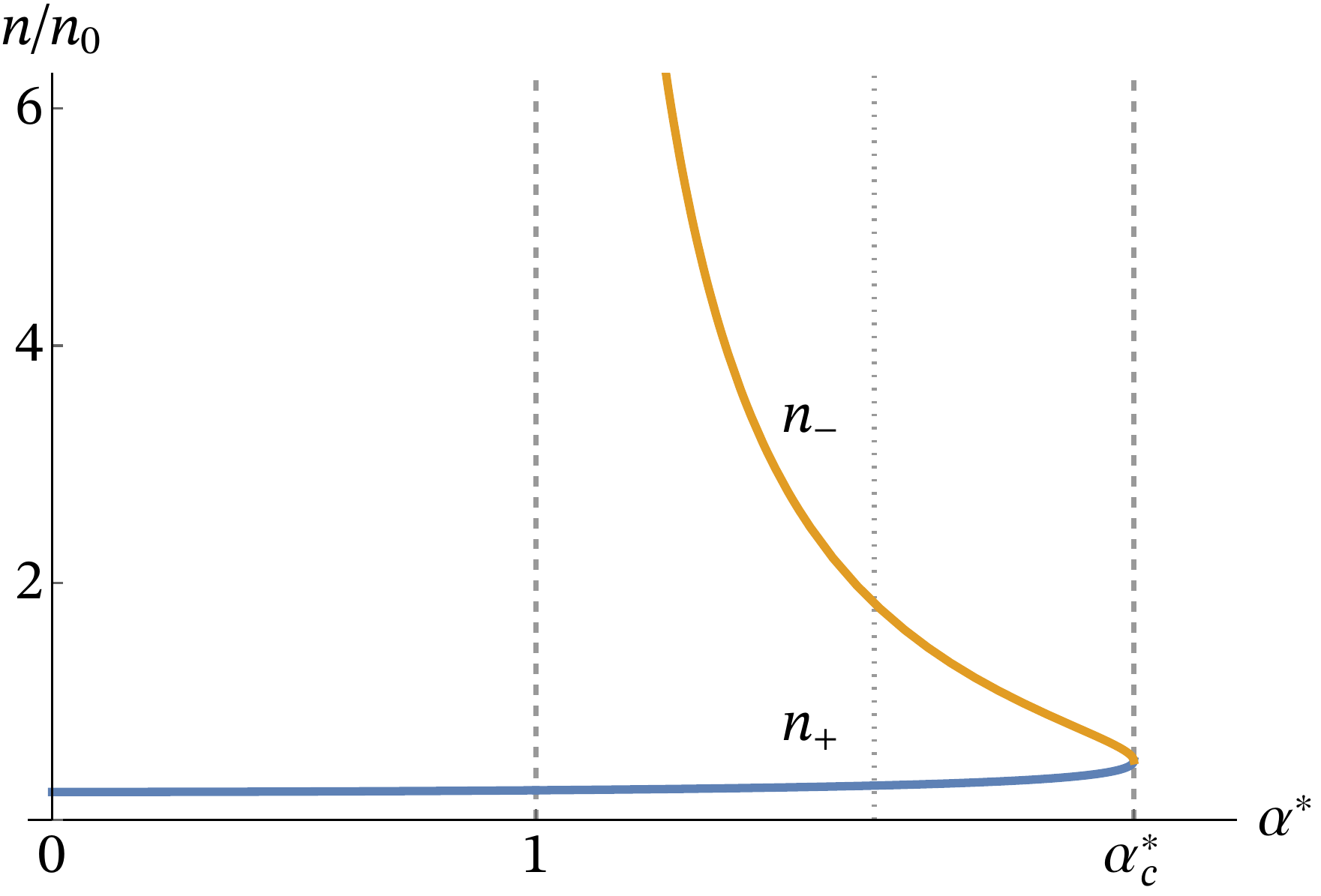}
	\caption{The electron density dependence on the iSOI parameter for $\mathfrak{v}^* = 2$.}
\label{fig2}
\end{figure}

The electron density exhibits an \textit{S}-type dependence on $\alpha^*$ as seen from Fig.~\ref{fig2}. At weak iSOI $\alpha^* < 1$, the solution is unique. In the range of $1 < \alpha^* < \alpha^*_c$ there appear two solutions, the stability of which should be examined. At $\alpha^* > \alpha^*_c$ the solution is at all absent within the simple model considered. The critical iSOI magnitude is given by 
\begin{equation}
\label{alphas}
	\alpha^*_c = \sqrt{1 + {\mathfrak{v}^*}^2}\,.	
\end{equation}

Such behavior of $n(\alpha^*)$ indicates a possible instability of the electron system at sufficiently strong iSOI $\alpha^* \in (1,\alpha^*_c)$ and a tendency for a radical transformation of the electron state at $\alpha^* > \alpha^*_c$, which may lead to the emergence of spatially inhomogeneous structures or a new correlated state. Nontrivial effects are expected already when $\alpha^*$ is of the order of unity. Our estimates show that such values of $\alpha^*$ can be attained in materials with a strong spin-orbit interaction~\cite{manchon2015new}. Presently the tunable RSOI with the parameter as large as $\alpha_R\sim 4\times 10^{-10}$~eV m is attained in such materials as Bi$_2$Se$_3$ in quantum wells in the presence of the electric field of the order of $3 \times 10^5$~V/cm~\cite{PhysRevLett.107.096802,manchon2015new}. Using these data one can estimate the distance $a$ between the electron system and the gate at which $\alpha^*\sim 1$. For $m=0.1 m_e$ and $\epsilon \sim 10$ we estimate $a\sim 40$~Å\@, which is realizable in modern heterostructures.  

Mechanisms stabilizing the electron system at strong iSOI and the nature of the emerging electron state constitute a challenging problem that deserves a separate study. A possible mechanism should include the processes leading to an essential rearrangement of the density of states, such as the population of the higher transverse sub-bands in the quantum wire and the formation of a new correlated state. 

\section{Collective modes}

In this section we study the spectra of collective excitations in a one-dimensional (1D) electron system below a threshold of a possible instability to find out the conditions under which the stability of the excitations could be lost.

An important aspect of the iSOI is a nontrivial modification of the electron-electron (e-e) interaction Hamiltonian. The image charges not only screen the Coulomb interaction to make it dipole-like, but also create a new spin-dependent component of the e-e interaction. This effect should be manifested in a qualitative change in the correlation functions. To the best of our knowledge, the properties of the correlated electron state and its collective excitations were not investigated in literature in such circumstances.

Our model Hamiltonian reads as
\begin{equation}
	H = H_\mathrm{kin} + H_\mathrm{e-e} + H_\mathrm{iSOI}\,.
\label{fullham}
\end{equation}
The first term is the kinetic energy $H_\mathrm{kin} = {(2m)}^{\!-1} \sum_{s} \int dx\, \psi^+_{s}(x)\mathfrak{p}_x^2 \psi_{s}(x)$, where  $\psi_{s}(x)$ stands for the electron field operator and $\mathfrak{p}_{x}$ stands for momentum. 

The operator of the e-e interaction energy is
\begin{equation}
\label{Hee}
	\begin{split}
		H_\mathrm{e-e} = &\frac{1}{2} \sum_{s_1 s_2} \int 
		 \psi^+_{s_1}(x_1) \psi^+_{s_2}(x_2) \mathcal{U}(x_1-x_2) \\ 
		&\! \times \psi_{s_2}(x_2) \psi_{s_1}(x_1)\,dx_1 dx_2\,.
	\end{split}
\end{equation}
Here $\mathcal{U}(x) = \frac{e^2}{\sqrt{x^2 + d^2}} - \frac{e^2}{\sqrt{x^2 + a^2}}$ is the e-e interaction potential screened by the image charges.
Its Fourier transform $U_q = \int dx\, \mathcal{U}(x) e^{-iqx}$ is a table integral~\cite{gradshteyn2014table}, equal to $U_q = 2e^2\left(K_0(qd) - K_0(qa)\right)$, with $K_0$ being the modified Bessel function~\cite{olver}.

The iSOI Hamiltonian can be formulated on the basis of the standard form~\eqref{rham} taking into account that the electric field is produced by all the charges in the system. Using Eq.~\eqref{rham} in the case of the iSOI is supported by calculations carried out in Ref.~\cite{0953-8984-16-39-017} within the relativistic multiple-scattering methods.

The iSOI Hamiltonian reads as
\begin{equation}
\label{soi_fq}
	H_\mathrm{iSOI} = \frac{\alpha}{\hbar} \sum_{i} \frac{1}{2}
	\left[\mathfrak{E}_y(x_i)\mathfrak{p}_{x_i} + \mathfrak{p}_{x_i}\mathfrak{E}_y(x_i)\right]\sigma_{z_i}\,,
\end{equation}
where $\sigma_{z_i}$ is the Pauli matrix of the $i$th electron and $\mathfrak{E}_y(x_i)$ is the $y$ component of the electric field acting on the electron. This field contains two principally different contributions that come from external charges and the images of all electrons in the system. We emphasize that the iSOI can not be described by a single-particle Hamiltonian as opposed to RSOI described in Refs.~\cite{PhysRevB.62.16900,hausler2001rashba,PhysRevB.66.075331,governale2002spin,PhysRevB.71.115322,PhysRevLett.98.126408,braunecker2010spin,PhysRevB.88.125143,PhysRevB.91.161305} by a fixed parameter $\alpha_R$.

The two-particle contribution is the total field of other electron images acting on a given electron,
\begin{equation}
	\mathfrak{E}_y^{ee}(x_i) = \sum_{j\ne i}\mathcal{E}(x_i - x_j)\,,
\end{equation}
where $\mathcal{E}(x_i - x_j) = -e a{\left[{(x_i - x_j)}^2 + a^2\right]}^{\! -(3/2)}$. A corresponding collective contribution to the Hamiltonian~\eqref{soi_fq} equals
\begin{equation}
\label{soi_sq}
	\begin{split}
		H_\mathrm{iSOI} = &\frac{\alpha}{2\hbar} \sum_{s_1s_2}  \int \psi^+_{s_1}(x_1) \psi^+_{s_2}(x_2) \left[ \mathcal{E}(x_1-x_2)\mathcal{S}_{12} \right. \\
		&{}+ \left. \mathcal{S}_{12} \mathcal{E}(x_1-x_2) \right] \psi_{s_2}(x_2) \psi_{s_1}(x_1)\,dx_1 dx_2\,,
	\end{split}
\end{equation} 
with $\mathcal{S}_{12} = (\mathfrak{p}_{x_1}s_1 + \mathfrak{p}_{x_2}s_2)/2$.

The Hamiltonian~\eqref{soi_sq} together with Eq.~\eqref{Hee} forms a modified Hamiltonian of the e-e interaction that contains a spin-dependent component appearing because of iSOI\@.

A single-particle contribution, coming from the image of the positive background charge $n_\mathrm{ion}$ in the wire (and the charge in the gate, should there be any) as well as the field of the electron's own image $\mathcal{E}(0)$ equals $\mathfrak{E}_y^{0} = \mathcal{E}(0) - n_\mathrm{ion}E_0$, where $E_0$ is the $q=0$ component of the Fourier-transform $E_q = -2e |q| K_1(|q|a)$ of the field $\mathcal{E}(x)$~\cite{gradshteyn2014table}. This leads to a single-particle contribution to the Hamiltonian~\eqref{soi_fq},
\begin{equation}
\label{soi0}
	H_\mathrm{iSOI}^{0} = \frac{\alpha}{\hbar}\sum_{s}\int dx\, 
	\psi^+_{s}(x)\mathfrak{E}_y^{0}\,\mathfrak{p}_{x} s \psi_{s}(x)\,.
\end{equation}

Below we investigate a linear response of the system defined by the Hamiltonian~\eqref{fullham}--\eqref{soi0} to an external perturbation of the form $H_\mathrm{ext} = \sum_{s} \int dx\, \psi^+_{s}(x) \varphi^{(s)}(x,t) \psi_{s}(x)$. The calculations are based on two independent methods, viz.\ the random phase approximation (RPA) and bosonization. Both approaches yield compatible results. The calculations are performed for a 1D system of length $L$ with fixed mean electron density $n_0$. The periodic boundary conditions are imposed, and the limit $L\to \infty$ considered. 

\subsection{RPA approach}

RPA calculations are based on the equation of motion for the Wigner function derived in the Appendix. The Fourier components $n^{(s)}_{q\omega}$ of electron density with the $z$ component of spin $s$, wave-vector $q$, and frequency $\omega$ are shown to satisfy the following system of linear equations:
\begin{align}
\label{linearsystem}
		n^{(s)}_{q\omega} &\left(\chi^{-1}_{q\omega} - U_q + \frac{m\alpha^2E_q}{\hbar^2}(2\mathcal{F}_0 + n_0E_q) - s\,\omega\frac{2m\alpha E_q}{\hbar q}\right) \notag \\
		&{} + n^{(-s)}_{q\omega}\left( - U_q + \frac{m\alpha^2E_q}{\hbar^2}(2\mathcal{F}_0 + n_0E_q)\right) = \varphi^{(s)}_{q\omega}\,.
\end{align}
The mean electric field $\mathcal{F}_0 = \mathcal{E}(0) + (n_0 - n_\mathrm{ion})E_0$ as compared to $\mathfrak{E}_y^{0}$ contains additionally the contribution from the mean electron density. By $\chi_{q\omega}$ we denote the Lindhard susceptibility,
\begin{equation}
	\chi_{q\omega} = \frac{m}{2\pi \hbar^2 q}\ln \frac{{(q - 2k_F)}^2 - {\left(\frac{2m\omega + i0}{\hbar q}\right)}^2}{{(q + 2k_F)}^2 - {\left(\frac{2m\omega + i0}{\hbar q}\right)}^2}\,,
\end{equation}
where $k_F = \pi n_0/2$.
\begin{figure}[t]
	\includegraphics[width=0.9\linewidth]{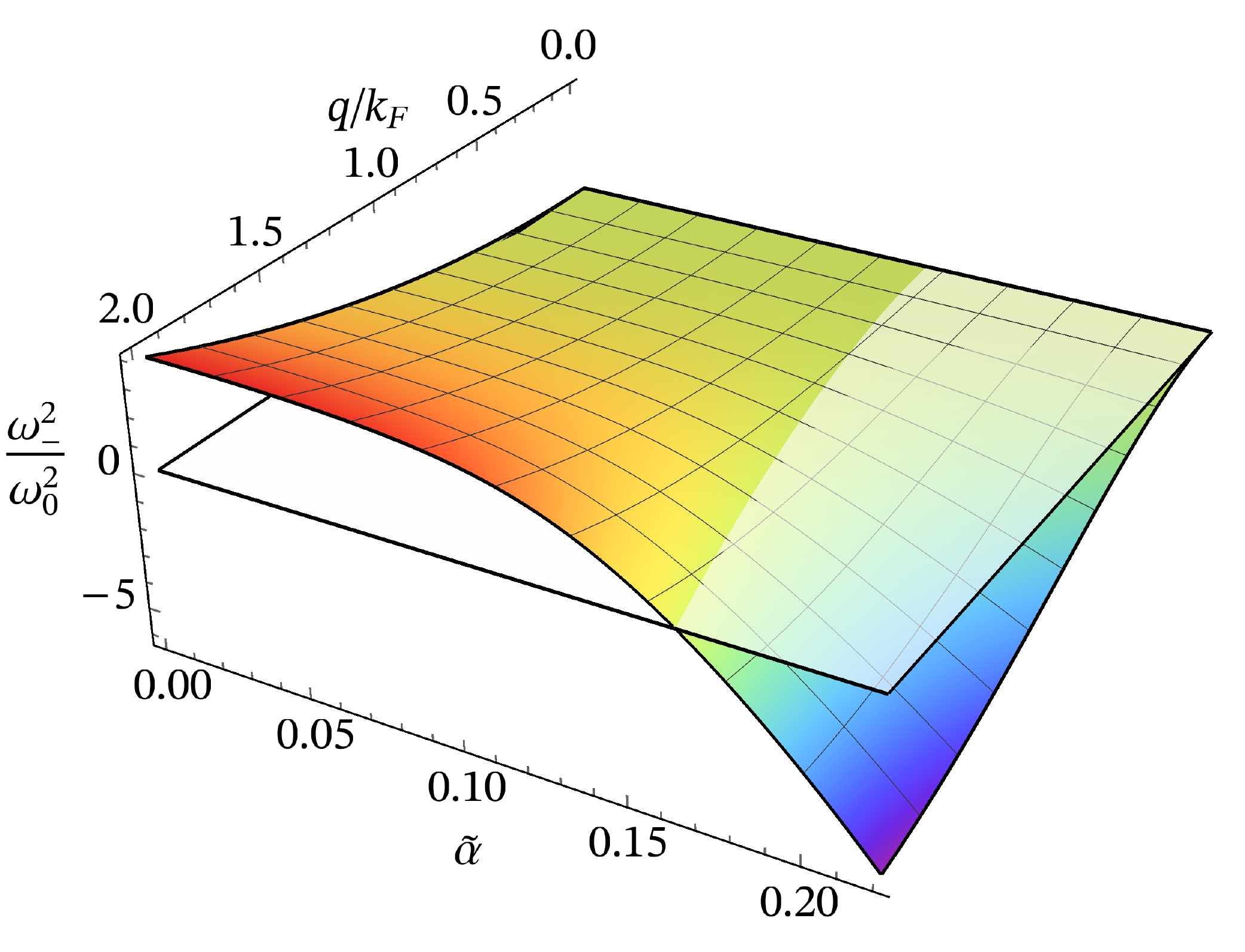}
	\caption{The square of the frequency $\omega_{-}^2$ of collective excitations as a function of wave vector and iSOI amplitude. Additionally, a plane $\omega_{-}^2 = 0$ is shown. The frequency is normalized at $\omega_0 = v_F k_F$. The system parameters are taken as follows: $k_Fa_B = 1.27$, $d = 0.078 a_B$, $a = 0.39 a_B$, $n_\mathrm{ion} = n_0$.}
\label{fig3}
\end{figure}

Setting the determinant of~\eqref{linearsystem} to zero, we obtain the dispersion equation for both branches of collective excitations,
\begin{equation}
\label{dispersion}
	\begin{split}
		{\left(\frac{\omega_{\pm}}{qv_F}\right)}^2 = 
		{}&1 + \left(\tilde{U}_q - \tilde{\alpha}^2\tilde{\mathcal{F}}_0\tilde{E}_q\right)\\
		&\pm \sqrt{{\left(\tilde{U}_q - \tilde{\alpha}^2\tilde{\mathcal{F}}_0\tilde{E}_q\right)}^2 + \tilde{\alpha}^2\tilde{E}_q^2}\,.
	\end{split}
\end{equation}
Dimensionless amplitudes are $\tilde{\alpha} = \dfrac{2}{\pi}\dfrac{\alpha n_0}{e a_B}$, $\tilde{U}_q = \dfrac{U_q}{\pi \hbar v_F}$, $\tilde{\mathcal{F}}_0 = \dfrac{\mathcal{F}_0}{e n_0^2}$, and $\tilde{E}_q = \dfrac{E_q}{e n_0}$ with $v_F = \dfrac{\hbar k_F}{m}$ and $a_B=\hbar^2/ m e^2$.

Of most interest is branch $\omega_{-}$ since it has an unusual dependence on the wave-vector $q$ and the iSOI parameter $\tilde{\alpha}$. This dependence is demonstrated in Fig.~\ref{fig3} in the case where the distance $a$ is small enough. The frequency of this mode and its velocity decrease with increasing $\tilde{\alpha}$. The frequency squared $\omega_{-}^2(q)$ turns to zero at some condition,
\begin{equation}
\label{alphat}
	\tilde{\alpha} = \tilde{\alpha}_q^0 \equiv \dfrac{\sqrt{1 + 2\tilde{U}_q}}{\sqrt{\tilde{E}_q^2 + 2\tilde{\mathcal{F}}_0\tilde{E}_q}}\,,
\end{equation}
and even becomes negative in the region of $\tilde{\alpha} > \tilde{\alpha}^0_q$, where the excitations become unstable. It is worth noting that upon the increase in $\tilde{\alpha}$ the excitations start losing their stability in the long-wave region where also the largest frequency increment appears in the instability regime.

\begin{figure}[t]
\includegraphics[width=0.9\linewidth]{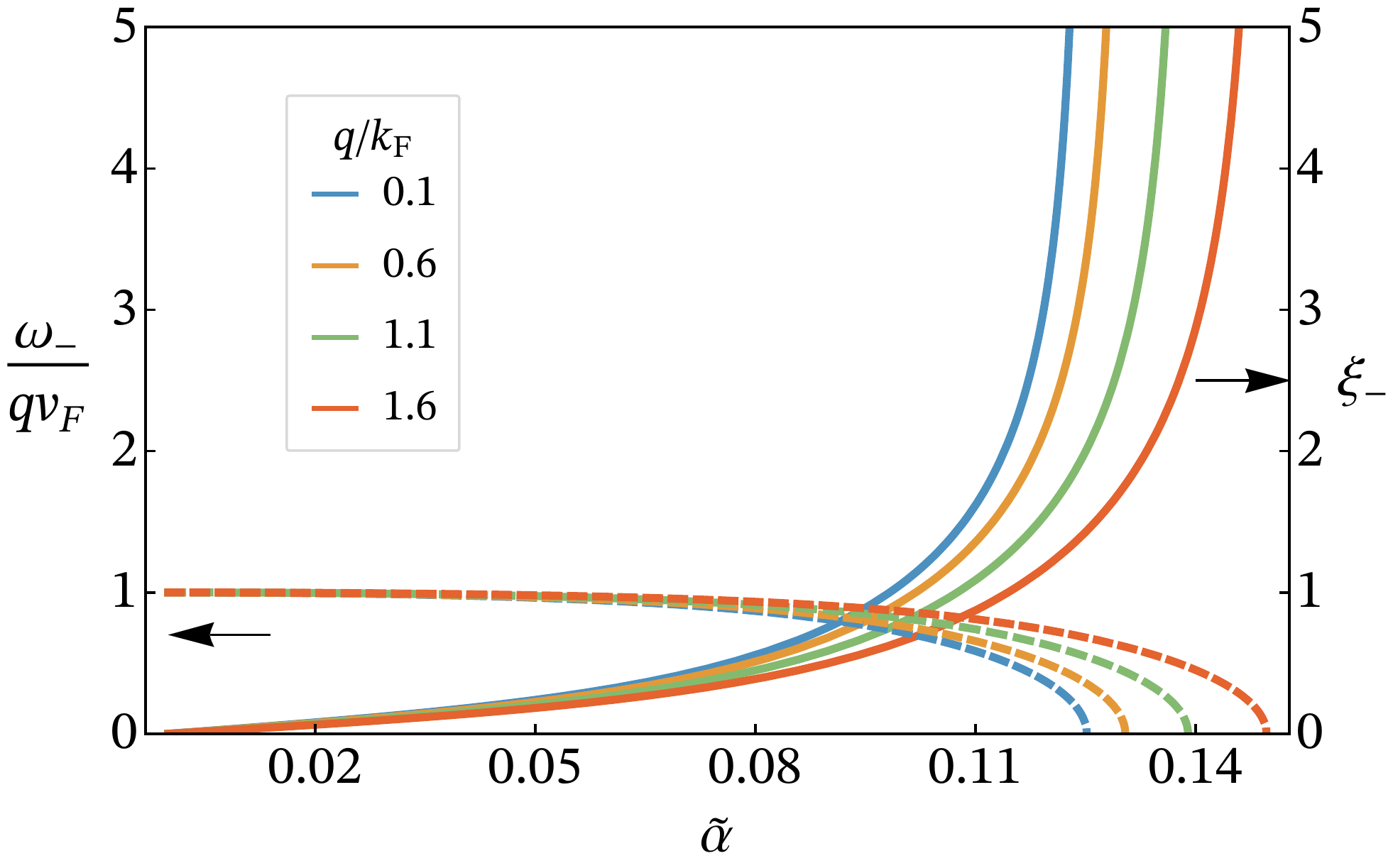}
	\caption{The spin-charge separation parameter (solid line) and normalized phase velocity (dashed line) for the $\omega_{-}$ collective mode as a function of iSOI amplitude. The same system parameters as in Fig.~\ref{fig3}.}
\label{fig4}
\end{figure}

Spin-dependent interactions break the spin-charge separation between the branches $\omega_{\pm}$ of collective  excitations. It is interesting to investigate how the spin-charge structure of the excitations evolves as $\tilde{\alpha}$ is increased. From Eq.~\eqref{linearsystem} we determine the spin-charge separation parameter $\xi_{\pm}$ for both branches of excitations,
\begin{equation}
	\xi_{\pm} = \frac{n^{+}_{q\omega} + n^{-}_{q\omega}}{n^{+}_{q\omega} - n^{-}_{q\omega}}\Bigg|_{\omega_{\pm}}\!\! = \frac{1}{\tilde{\alpha}\tilde{E}_q}\left(\frac{\omega_{\pm}}{q v_F} - \frac{q v_F}{\omega_{\pm}}\right).
\end{equation}
At $\tilde{\alpha} = 0$, the parameter $\xi_{-} = 0$, which means that branch $\omega_{-}$ corresponds to a purely spin excitation ($n^{+}_{q\omega} = -n^{-}_{q\omega}$) with dispersion law $\omega_{-} = v_F q$. However, as $\tilde{\alpha} \to \tilde{\alpha}^0_q$, the frequency $\omega_{-}(q) \to 0$ and the parameter $\xi_{-} \to \infty$ as shown in Fig.~\ref{fig4}. Consequently, near the threshold $\tilde{\alpha} = \tilde{\alpha}^0_q$ the collective excitation $\omega_{-}(q)$ turns into a purely charge excitation ($n^{+}_{q\omega} = n^{-}_{q\omega}$). 

The system stiffness $\varkappa = - \lim_{q \to 0} \chi_{nn}^{-1}(q,0)$ with a charge susceptibility $\chi_{nn}(q,\omega) = (n^{+}_{q\omega} + n^{-}_{q\omega})/\varphi_{q\omega}$ determined from Eq.~\eqref{linearsystem} equals
\begin{equation}
\label{stiffness}
	\varkappa = \pi\hbar v_F(1 + 2\tilde{U}_{0})\left[1 - {\left(\frac{\tilde{\alpha}}{\tilde{\alpha}^0_0}\right)}^2\right]\;.
\end{equation}
The stiffness turns to zero at $\tilde{\alpha} = \tilde{\alpha}^0_0$. This points at the instability of the charge subsystem. This is the most pronounced manifestation of the iSOI in the e-e correlations.

\begin{figure}[t]
	\includegraphics[width=0.9\linewidth]{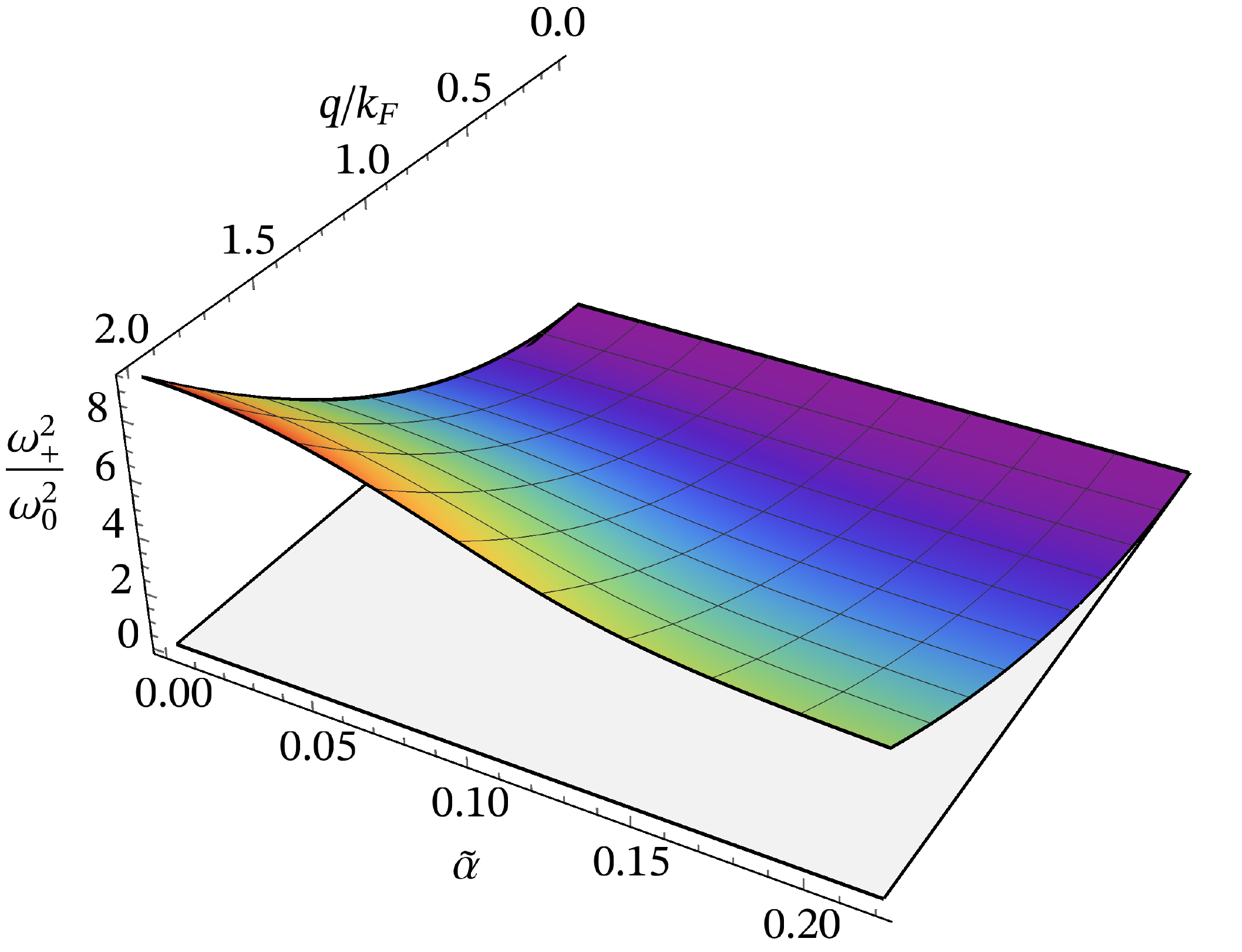}
	\caption{The square of the frequency $\omega_{+}^2$ of collective excitations as a function of wave vector and iSOI amplitude. Additionally, a plane $\omega_{+}^2 = 0$ is shown. The same system parameters as in Fig.~\ref{fig3}.}
\label{fig5}
\end{figure}

On the contrary, at $\tilde{\alpha} = 0$ another branch $\omega_{+}$ corresponds to purely charge excitations, which transform into purely spin ones as $\tilde{\alpha}$ increases. Their spectrum is shown in Fig.~\ref{fig5}. Upon the increase in $\tilde{\alpha}$ their velocity $v_{+}(q)$ always remains positive. The stiffness of the spin subsystem does not turn to zero.

Let us compare the critical iSOI value $\tilde{\alpha}^0_0$ from Eq.~\eqref{alphat} at which the long-wave collective excitations start losing their stability with $\alpha^*_c$ of Eq.~\eqref{alphas}, corresponding to the instability of the ground state of a system with
a fixed chemical potential. For the case when the system is sufficiently close to the gate $n_0 a \ll 1$, we obtain for dimensional iSOI values $\tilde{\alpha}^0_0 \propto \sqrt{n_0 a} \, \alpha^*_c$, which means that the collective excitations instability develops first. In the opposite limiting case of $n_0 a \gg 1$, they are of the same order of magnitude.

\subsection{Bosonization approach}

The bosonization~\cite{haldane1981luttinger} treatment of the problem leads to similar results. The presentation is simplified greatly in the absence of the mean electric field $\mathcal{F}_0$. Then the eigenstates of the kinetic energy can be chosen as the basis functions. Linearizing their spectrum, we introduce the bosons $a_s^+(q) = {(\frac{2\pi}{L|q|})}^{1/2}\sum\limits_{r = \pm} \theta(rq)\rho_{rs}(q)$ where the normal ordered density of fermions with spin projection $s$ on branch $r$ is $\rho_{rs}(q) = \sum_p : c^+_{rs}(p+q)c_{rs}(p)\colon $. The quadratic part of a bosonized Hamiltonian~\eqref{fullham} is
\begin{align}
\label{bosham}
	H = {}&\frac{\hbar v_F}{2} \sum \limits_{\substack{q>0\\r,s=\pm}} q 
	\left[ a_s^+(rq) a_s(rq)(2 + \tilde{U}_q + rs\tilde{\alpha}\tilde{E}_q)\right. \notag \\
	&+ \frac12[a_s^+(rq) a_{-s}^+(-rq) + H.c.](\tilde{U}_q + rs\tilde{\alpha}\tilde{E}_q) \\
	&+ \left. \frac12[a_s^+(rq)a_{s}^+(-rq) + a_s^+(rq)a_{-s}(rq) + H.c.]\tilde{U}_q \right] \notag\,.
\end{align}
We diagonalize it by the Bogoliubov-Tyablikov transformation~\cite{avery1976creation}. For this purpose, matrices defining commutators $[H,a_k^+] = \sum_i a_i^+ A_{ik} + a_i B_{ik}$ for each boson $a_k$ from~\eqref{bosham} are constructed. Then the squares of the elementary excitations frequencies are just the eigenvalues of matrix $(A - B)(A + B)$. They are
\begin{equation}
	{\left(\frac{\omega_{\pm}}{qv_F}\right)}^2 = 1 + \tilde{U}_q \pm \sqrt{\tilde{U}_q^2 + \tilde{\alpha}^2\tilde{E}_q^2}\;,
\end{equation}
which coincides with~\eqref{dispersion} at $\mathcal{F}_0 = 0$.

\section{Conclusion}
In conclusion, we have shown that the Coulomb interaction of 1D electrons with the image charges in the nearby metallic gate has a spin-dependent component caused by the Rashba spin-orbit interaction. This iSOI can strongly affect both the ground state of the system and the collective excitations. The main effect is an instability which occurs as the iSOI parameter is large enough. Our estimations have shown that the critical conditions are attainable in realistic systems. This effect seems to be rather general for a wide class of 1D, quasi-1D, and 2D systems in materials with strong spin-orbit interaction. The instability leads to the formation of a new correlated state that needs to be investigated further.

\section*{Acknowledgments}
This work was partially supported by the Russian Foundation for Basic Research (Grant No 17--02--00309).

\onecolumngrid
\appendix
\section{}
Here we derive Eq.~\eqref{linearsystem} of the main text. Define single- and two-particle Klimontovich operators~\cite{klimontovich} as
\begin{equation}
	\hat{f}^{(s)}(x,p,t) = \frac{1}{2 \pi} \int d\eta\, e^{i p\eta}\psi_s^+(x + \frac{\eta}{2},t)\psi_s(x - \frac{\eta}{2},t)
\end{equation}
and
\begin{equation}
	\hat{f}^{(s_1, s_2)}(x_1,p_1,x_2,p_2,t) = \frac{1}{{(2 \pi)}^2} \int d\eta_1 d\eta_2\, e^{i (p_1 \eta_1 + p_2 \eta_2)} \psi_{s_1}^+(x_1 + \frac{\eta_1}{2},t)\psi_{s_2}^+(x_2 + \frac{\eta_2}{2},t)\psi_{s_2}(x_2 - \frac{\eta_2}{2},t)\psi_{s_1}(x_1 - \frac{\eta_1}{2},t)\,.
\end{equation}

The average values $f^{(s)}(x,p,t)$ and $f^{(s_1 s_2)}(x_1,p_1,x_2,p_2,t)$ of these operators w.r.t.\ the ground state are just the Wigner distribution functions (WDFs), which allow one to find the observables of interest. Thus, the electron density is expressed as
\begin{equation}
	n^{(s)}(x,t) = \int dp\, f^{(s)}(x,p,t)\,.
\end{equation}

By commuting $\hat{f}^{(s)}(x,p,t)$ with $H+H_\mathrm{ext}$ and taking the average, the equation of motion for the WDF is obtained,
\begin{align}
\label{BBGKY}
	i \hbar \partial_t f^{(s)}(x,p,t) = &- \frac{i \hbar^2 p}{m} \partial_x f^{(s)}(x,p,t) + \frac{1}{2\pi} \int d\eta dp_1\, e^{i(p - p_1)\eta} f^{(s)}(x,p_1,t)\left(\varphi_s(x - \frac{\eta}{2},t) - \varphi_s (x + \frac{\eta}{2},t)\right)\\
	&{}+ \frac{1}{2\pi} \sum_{\varsigma}\int d\xi d\eta\, dp_1 dp_2\, e^{i(p - p_1)\eta} f^{(s, \varsigma)}(x,p_1,\xi,p_2,t) \left(\mathcal{U} (x - \xi - \frac{\eta}{2}) - \mathcal{U} (x - \xi + \frac{\eta}{2})\right) \notag \\ 
	&{}- i \alpha s \mathfrak{E}_y^0 \, \partial_x f^{(s)}(x,p,t) - \frac{i \alpha s}{2\pi}\int d\xi d\eta\, dp_1 dp_2\, e^{i(p - p_1)\eta} f^{(s, -s)}(x,p_1,\xi,p_2,t) \left(\mathcal{E}' (x - \xi - \frac{\eta}{2}) + \mathcal{E}' (x - \xi + \frac{\eta}{2})\right) \notag\\
	&{}- \frac{i \alpha}{2\pi} \sum_{\varsigma} \varsigma \int d\xi d\eta\, dp_1 dp_2\, e^{i(p - p_1)\eta} \left( \frac12 \partial_{\xi}f^{(s, \varsigma)}(x,p_1,\xi,p_2,t) + i p_2f^{(s, \varsigma)}(x,p_1,\xi,p_2,t) \right) \mathcal{E} (x - \xi - \frac{\eta}{2}) \notag \\
	&{}- \frac{i \alpha}{2\pi} \sum_{\varsigma} s \int d\xi d\eta\, dp_1 dp_2\, e^{i(p - p_1)\eta} \left( \frac12 \partial_{x}f^{(s, \varsigma)}(x,p_1,\xi,p_2,t) + i p_1f^{(s, \varsigma)}(x,p_1,\xi,p_2,t) \right) \mathcal{E} (x - \xi - \frac{\eta}{2}) \notag \\
	&{}- \frac{i \alpha}{2\pi} \sum_{\varsigma} \varsigma \int d\xi d\eta\, dp_1 dp_2\, e^{i(p - p_1)\eta} \left( \frac12 \partial_{\xi}f^{(s, \varsigma)}(x,p_1,\xi,p_2,t) - i p_2f^{(s, \varsigma)}(x,p_1,\xi,p_2,t) \right) \mathcal{E} (x - \xi + \frac{\eta}{2}) \notag \\
	&{}- \frac{i \alpha}{2\pi} \sum_{\varsigma} s \int d\xi d\eta\, dp_1 dp_2\, e^{i(p - p_1)\eta} \left( \frac12 \partial_{x}f^{(s, \varsigma)}(x,p_1,\xi,p_2,t) - i p_1f^{(s, \varsigma)}(x,p_1,\xi,p_2,t) \right) \mathcal{E} (x - \xi + \frac{\eta}{2}) \notag\,.
\end{align}
This is the first equation in the Bogoliubov-Born-Green-Kirkwood-Yvon hierarchy~\cite{bonitz2016quantum}. We truncate it using the RPA by factorizing the two-particle WDF~\cite{hasegawa1975electron},
\begin{equation}
	f^{(s_1, s_2)}(x_1,p_1,x_2,p_2,t) = f^{(s_1)}(x_1,p_1,t)f^{(s_2)}(x_2,p_2,t)\,.
\end{equation}
This defines the way the pair correlations are taken into account. Introduce the deviation $f^{(s)}_1(x,p,t)$ of $f^{(s)}(x,p,t)$ from its equilibrium value $f_0^{(s)}(p)$ as a result of the external perturbation $H_\mathrm{ext}$,
\begin{equation}
	f^{(s)}_1(x,p,t) = f^{(s)}(x,p,t) - f^{(s)}_0(p)\,.
\end{equation}

\newpage
\twocolumngrid

The equation of motion for $f^{(s)}_1(x,p,t)$, linearized w.r.t. $H_\mathrm{ext}$, in Fourier representation reads as
\begin{align}
	\label{eqmot1}
	&- \hbar\omega f^{(s)}_1(q,p,\omega) = -\frac{\hbar^2 pq}{m}f^{(s)}_1(q,p,\omega)\\
	\label{eqmot2}
	&{}+ \varphi^{(s)}_{q\omega}\left[f^{(s)}_0(p+\frac{q}{2}) - f^{(s)}_0(p-\frac{q}{2})\right]\\
	\label{eqmot3}
	&{}+ U_q\left[f^{(s)}_0(p+\frac{q}{2})-f^{(s)}_0(p - \frac{q}{2})\right]\sum_{\varsigma} n^{(\varsigma)}_{q\omega}\\
	\label{eqmot4}
	&{}- \alpha qs\,\mathcal{F}_0 f^{(s)}_1(q,p,\omega)\\
	\label{eqmot5}
	&{}+ \alpha s E_q p\left[f^{(s)}_0(p + \frac{q}{2}) - f^{(s)}_0(p - \frac{q}{2})\right]\sum_{\varsigma} n^{(\varsigma)}_{q\omega}\\
	\label{eqmot6}
	&{}+ \alpha E_q \left[f^{(s)}_0(p + \frac{q}{2}) - f^{(s)}_0(p - \frac{q}{2})\right]\sum_{\varsigma}\varsigma \int \kappa f^{(\varsigma)}_1(q,\kappa,\omega)\,d\kappa\,.
\end{align}

\vspace{-0.27cm}
The terms~\eqref{eqmot1}--\eqref{eqmot3} reflect the contribution of kinetic energy, external potential and Coulomb e-e interaction.

The term~\eqref{eqmot4} reflects the part of iSOI due to the mean electric field $\mathcal{F}_0 = \mathcal{E}(0) + (n_0 - n_\mathrm{ion})E_0$. Let us discuss the effects of the mean field in some more detail. RPA assumes that the single-particle states, the distribution over which is given by $f^{(s)}_0(p)$, are formed by a single-particle part of the Hamiltonian, the mean electric field included. For a system with a fixed particle number, this sets the Fermi momenta for a spin direction $s$ to be $k_F^{(s)} = -s k_\mathrm{so} \pm k_F$, where $k_\mathrm{so} = \alpha m \mathcal{F}_0/\hbar^2$ and $k_F$ stands for $\pi n_0/2$. Restricting the equation of motion to include just terms~\eqref{eqmot1}--\eqref{eqmot4}, we easily can find the electron density for the case of iSOI exclusively due to the mean field. For this purpose express the $f^{(s)}_1(x,p,t)$ and integrate over $p$ to obtain the equations for the density,
\begin{equation}
\label{meanfield}
	n_{q \omega}^{(s)} = \varphi_{q\omega}^{(s)}\chi_{q\omega}^{(s)} + U_q \chi_{q\omega}^{(s)} \sum_{\varsigma} n_{q \omega}^{(\varsigma)}\,.
\end{equation}
Here the Lindhard susceptibility
\begin{equation}
	\begin{split}
		\chi^{(s)}_{q\omega} &= \int d\kappa\, \frac{f^{(s)}_0(\kappa + \frac{q}{2}) - f^{(s)}_0(\kappa - \frac{q}{2})}{-\hbar (\omega+i0) + \frac{\hbar^2 \kappa q}{m} + \alpha qs\,\mathcal{F}_0}\\
		&= \frac{m}{2\pi \hbar^2 q}\ln \frac{{(q - 2k_F)}^2 - {\left(\frac{2m\omega + i0}{\hbar q}\right)}^2}{{(q + 2k_F)}^2 - {\left(\frac{2m\omega + i0}{\hbar q}\right)}^2}
	\end{split}
\end{equation}
turns out to be independent of spin $s$ and of the mean-field $\mathcal{F}_0$. Hence, the collective excitations, the dispersion relation of which
\begin{equation}
	\chi^{-1}_{q\omega}[\chi^{-1}_{q\omega}-2U_q] = 0
\end{equation}
is obtained by setting the determinant of~\eqref{meanfield} to zero, are the spin-charge separated common plasmons and spinons. Their velocity does not depend on SOI\@.

The terms~\eqref{eqmot5} and~\eqref{eqmot6} reflect the collective electron contribution to iSOI\@. Whereas the structure of the term~\eqref{eqmot5} resembles the Coulomb contribution~\eqref{eqmot3}, there also appears a qualitatively new integral term~\eqref{eqmot6}. Integrate the equation of motion w.r.t. $p$ to get
\begin{equation}
\label{intterm}
	\begin{split}
		(-\hbar \omega + \alpha q s \mathcal{F}_0)n_{q \omega}^{(s)} = 
		&- \frac{\hbar^2 q}{m} \int \kappa f^{(s)}_1(q,\kappa,\omega)\,d\kappa \\
		&{}- \alpha q s E_q \frac{n_0}{2} \sum_{\varsigma} n_{q \omega}^{(\varsigma)}\,.
	\end{split}
\end{equation}
Substitute the integral term from Eq.~\eqref{intterm} to Eq.~\eqref{eqmot6}, express $f^{(s)}_1(q,p,\omega)$, and integrate the latter w.r.t.\ $p$ to obtain the Eq.~\eqref{linearsystem} of the main text.

\bibliography{paper}

\end{document}